"EBK" : Leveraging Crowd-Sourced Social Media Data to Quantify How Hyperlocal Gang Affiliations Shape Personal Networks and Violence in Chicago's Contemporary Southside

Riley Tucker, Nakwon Rim, Alfred Chao, Elizabeth Gaillard, Marc G. Berman

## ABSTRACT

Recent ethnographic research reveals that gang dynamics in Chicago's Southside have evolved with decentralized micro-gang "set" factions and cross-gang interpersonal networks marking the contemporary landscape. However, standard police datasets lack the depth to analyze gang violence with such granularity. To address this, we employed a natural language processing strategy to analyze text from a Chicago gangs message board. By identifying proper nouns, probabilistically linking them to gang sets, and assuming social connections among names mentioned together, we created a social network dataset of 271 individuals across 11 gang sets. Using Louvain community detection, we found that these individuals often connect with gang-affiliated peers from various gang sets that are physically proximal. Hierarchical logistic regression revealed that individuals with ties to homicide victims and central positions in the overall gang network were at increased risk of victimization, regardless of gang affiliation. This research demonstrates that utilizing crowd-sourced, online information can enable the study of otherwise inaccessible topics and populations.

Research has shown that gangs and gang violence are serious issues facing society. Involvement in gangs heightens the likelihood that youth will engage in criminal behavior (Krohn & Thornberry, 2008), be victimized by gun violence (Papachristos , Braga, et al., 2015), become teenage parents (Thornberry et al., 1997), and drop out of school (Pyrooz, 2014). There is also evidence that the majority of urban youth violence may be gang-perpetuated (Kennedy et al., 2017), suggesting there are broad consequences when gangs proliferate. Interventions that could reduce the social harms associated with gangs are complicated by the fact that gangs have historically been challenging to access and research, with many studies relying on ethnographic methods that provide rich information at the cost of small scales and low generalizability (see: Pattillo, 1998; Stuart, 2020; Venkatesh, 2008).

To quantitatively study gangs at larger scales, scholars have analyzed records from police arrests and investigations. By connecting individuals who have been arrested or investigated together, scholars have measured criminal networks spanning entire neighborhoods and cities (Green et al., 2017; Papachristos et al., 2012, 2013). These works have shown that criminal network-based measures, such as network distance to homicide victims and the percentage of an individual's network that is gang affiliated, can indicate when an individual is at risk of dying to gun violence, offering important information that can guide the allocation of violence intervention resources. However, the data supporting these studies are notably limited in that they rely on bias-prone policing data that only capture relationships among people who are arrested or investigated together while excluding other types of relationships. As such, gang scholarship can be enhanced by developing alternative datasets that capture different dimensions of relationships among gang-affiliated people. While research has demonstrated that gang activity and discussion of gangs increasingly takes place online (Patton et al., 2017; Stuart, 2020; Stuart et al., 2020), scholars have yet to consider whether social media content could provide an alternative or supplementary window to view the social connectivity among gang-affiliated individuals. Such an approach is warranted, as scholars have shown we can learn social facts by tapping into crowd-sourced wisdom (Cooke, 1991; Davis-Stober et al., 2014; Surowiecki, 2005). To apply this approach to the context of gang scholarship, this study presents and validates a novel, computationally-driven strategy for constructing crowd-sourced, social-media-based gang networks. Using this approach, we demonstrate that gang-affiliated people are frequently socially connected to people from other gangs and that features derived from crowd-sourced social media networks can statistically explain mortality among gang-affiliated people.

*Police-Based Criminal Networks*

Contemporary gang scholars frequently examine the connectivity between gang-affiliated people in order to understand the dynamics leading to gang violence. At one analytic scale, scholars have identified alliances and conflicts between factions to better understand the nature of cross-group conflict. By using Chicago Police Department data that labeled homicide perpetrators and victims based on their gang affiliation, Papachristos (2009) generated a network capturing the frequency with which each gang perpetuated violence against other groups. Analysis of this network suggested that gang violence can be explained as a process of dominance disputes, where factions are more likely to engage in violent conflict when they have overlapping turf or when affiliates belong to the same racial group, with the effects of prior conflicts becoming "contagious" over time. In a follow-up study using comparable data from both Chicago and Boston, Papachristos and colleagues (2013) found that gangs are more likely to be connected through violent conflict when they are spatially proximate and when there is a history of conflict between groups. The authors argue that these results suggest a process of reciprocity where gangs will perpetuate violence in response to being victimized. More recent studies using sophisticated network methods have generated further evidence from multiple cities indicating that network dynamics among gangs drive violence in complex and nuanced ways (Bichler et al., 2020; Gravel et al., 2023; Lewis & Papachristos, 2020; Nakamura et al., 2020).

In response to this work situating gang scholarship within the network perspective, researchers have innovated beyond gang networks to measure the social networks of gang-affiliated people at the individual level. To create a social network approximating the connections among crime-involved people, Papachristos, Braga, and Hureau (2012) processed Field Intelligence Observation cards provided by the Boston Police Department to assign a network tie between any two individuals who were observed at the same time and location by police. By reviewing arrest records of verified gang members, researchers generated a network of 763 people who were connected to gang-affiliated individuals within one Boston community. Analysis of this network suggested that risk of victimization decreases with the network distance needed to connect an individual to a victim of gun violence, suggesting crime and victimization are contagious across individuals. Follow-up studies that leveraged arrest data to create larger "co-arrest" networks in Chicago and Newark showed that network effects are associated with gun violence across cities (Green et al., 2017; Papachristos, Braga, et al., 2015; Papachristos, Wildeman, et al., 2015; Papachristos & Wildeman, 2014). The literature is clear: people are at risk of victimization if other people in their criminal network get shot.

While scholars have used co-arrest data to generate strong evidence that network dynamics can explain violence, there are several limitations associated with co-offending networks generated from police arrests and investigations. First, this strategy of network measurement only captures a specific segment of any given individual's personal network; such networks likely fail to capture all of an individual's criminal connections and certainly do not represent all of a person's personal connections. Second, police have faced criticism for the accuracy of their data. For example, the Chicago Police Department's oft-criticized Strategic Subjects List is an implausibly large list of nearly 400,000 "high-risk" individuals that includes more than half of all African American men in Chicago in their twenties (Aspholm, 2020). If databases generated by police departments are error-prone or limited in their scope, analyses conducted on police data may also be limited or misrepresent how violence spreads through criminal networks. As such, we propose that developing alternative strategies to measure the social networks of criminal affiliates without the use of police data will be valuable for accurately and precisely measuring crime-related network dynamics. To this end, we propose one potential alternative utilizing a source of information that has only recently emerged: social media data.

*Social (Media) Networks*
Historically, there have been few sources outside of police databases offering substantive information about gangs and their affiliates. However, with the proliferation of the internet and social media, scholars have observed that gang conflicts are increasingly taking place online on the "digital streets" (Lane, 2018; Patton et al., 2017; Stuart, 2020). Shared spaces created through social media platforms offer opportunities for gang-affiliated people to make posts that symbolically signal their gang allegiances and 'diss' individuals affiliated with rival gangs (Stuart, 2020). Given this emerging phenomenon, scholars have begun to investigate how gang-affiliated people utilize social media services such as Twitter (Patton et al., 2017)**,** Facebook (Leverso & Hsiao, 2021), and YouTube (Lauger & Densley, 2018) to understand whether internet access has fundamentally changed gang interactions or simply created new venues for traditional interactions.

Despite the richness of information offered through online posts, only a small number of studies have attempted to study the networks of urban gangs using social media content. Leverso and Hsiao (2021) scraped comments and meta-data from a Facebook group about Chicago Latina/o gangs to quantify how often members of each gang sent negative comments to each other. Their results indicated that gangs were more likely to trade 'disses' when their turfs were geographically close to one another, suggesting a relationship between real-world interpersonal dynamics and online conflict. In a follow-up study that connected the Facebook-generated conflict network to Chicago Police Department shootings data, Hsiao, Leverso, and Papachristos (2023) demonstrated that gangs were more likely to have negative interactions online when there was a history of real-world violence between the groups. While these studies provide strong evidence that gang-related online interactions are embedded with contextual information from the real-world, to our knowledge, no studies have attempted to measure individual-level networks through such content, which is likely due to technological limitations. However, we propose that currently available, out-of-the-box natural language processing technologies can be used to analyze social media-derived data about gang networks and affiliations to provide new insights into the social forces driving violence among gang-affiliated people in the contemporary era.

*"EBK" and Gang Set Dynamics in Contemporary Chicago*
By utilizing rich, online information about gang-affiliated people, we can measure aspects of gangs that are generally inaccessible through traditional quantitative sources. Based on ethnographic evidence, Aspholm (2020) argues that African American gangs on Chicago's Southside have become decentralized, with large gang "nations" from the 1990s breaking into smaller factions of "sets" whose turfs may occupy just several city blocks. Aspholm suggests that, as a result of these changing dynamics, violence among today's gang-affiliated people in Chicago is not the product of conflict between gangs, as the literature generally assumes, but is instead the result of inter-personal conflicts among individuals. In summarizing the argument, Aspholm (2020) states, "...the cities traditional People and Folks coalitions have been effectively obsolete on the streets for nearly three decades. In short, today's gangs are not stable, hierarchical organizations with identifiable leaders and clear lines of authority. Gang animosities are fractured and diffuse, and violence is often erratic and unpredictable" (p. 149). Aspholm conceptualizes gang violence as an occurrence that can take place between any two people, even affiliates of the same gang. This framing, referred to as "EBK" (standing for "everybody killer" in Chicago gang parlance) dynamics, represents a strong theoretical departure from the traditional perspective which treats gang violence as the product of competition between factions.

To our knowledge, quantitative research on Chicago gangs has yet to consider or account for these smaller gang set factions that have been observed as a new norm in today's Chicago. Notably, all of the many studies of violence networks in Chicago analyze gangs at the larger nation level rather than at the set level despite Aspholm's (2020) argument that nation affiliations have lost meaning among today's gang-affiliated people. This is likely a result of limitations associated with reliance on Chicago Police Department data; our analysis of CPD fatal and non-fatal shooting arrests data found that only 47% of arrested individuals with a known nation affiliation had a known set affiliation. In other words, the majority of gang-affiliated people arrested for shootings did not have a set affiliation identified (53%). Notably, of the 2,395 shooting arrests with set labels, there were zero instances where the nation label was missing, providing further evidence that CPD focuses on labeling gang-affiliated people based on their nation while neglecting their set. As such, while qualitative research suggests that gang set dynamics have changed the processes leading to gang violence in Southside Chicago, the official police-based data that dominate the scholarly literature on gangs seemingly fail to capture gang affiliation at this hyper-localized scale (i.e., the gang set). This is extremely problematic if localized set affiliations are the most informative grouping variable for these individuals.

We propose that social-media generated networks are an alternative source of data that can be used to investigate socio-behavioral dynamics at a scale better suited to the modern gang landscape. Generally, scholars have shown that people utilize social media platforms to symbolize and reinforce their identities (Greenhow & Robelia, 2009). This holds true among gang-affiliated people, who frequently use social media content to highlight their allegiances and gang identities (Leverso & Hsiao, 2021; Stuart, 2020). If social media posts written by gang-affiliated people present the affiliations they find most authentic and meaningful, measuring gang affiliations through the lens of their social media followers should capture the level of affiliations that are most central in driving gang-related behavior and perpetuating violence. As such, using crowd-sourced social media data to study the geographic and social networks of gang-affiliated people can provide never-before-seen insights into the dynamics underpinning gang violence. To wit, this study is the first to quantitatively explore hyperlocal gang affiliations to assess how social network ties form and facilitate violence across Chicago gang sets.

**RESULTS**
*Construction of Social-Media Based Gang Network*
To measure networks of gang-affiliated people, this study draws from an online message board about Chicago gangs. **Figure 1** depicts the informational pipeline this study draws from. First, gang-affiliated people post content on social media. Online consumers of this content then develop canons about the friendships, rivalries, and acts of violence among relevant actors. These canons are then discussed on social media forums. We argue that these online discussions represent a process of crowd-sourced synthesis, where users communicate personal understandings of Chicago gangs and related events based on social media content to collectively develop a shared body of knowledge on this topic. By analyzing the synthesized content from such online spaces, we can quantify the accumulated knowledge about gangs for scientific purposes.

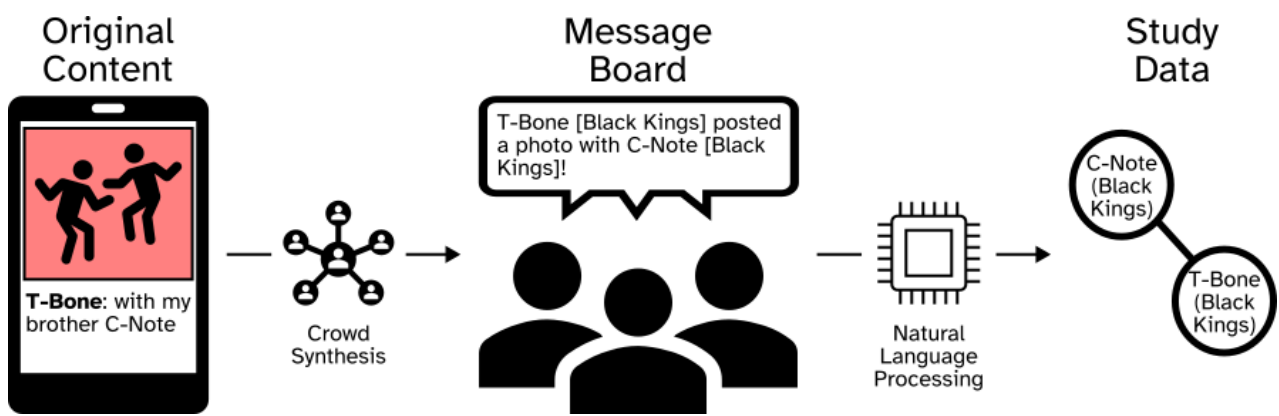

**Figure 1**. Conceptual diagram of data etiology. Gang-affiliated people post social media content emphasizing gang affiliations and activities. Online consumers process this information, engage in fandom, and discuss gang-related social media content in secondary online spaces (e.g., Reddit forums). By analyzing text from these secondary online spaces, we can learn about the social dynamics among gang-affiliated people.

Crowd-Sourced Gangs 6

We took advantage of a systematic posting style to extract aliases and gang affiliations from the text of 57.962 forum posts (See **Figure 2A**). Using parts-of-speech tagging, we identified 1,755 proper nouns that forum users had associated with a gang. We then used a probabilistic approach to link these named individuals to their most likely gang affiliation. By assuming that an individual was affiliated with a particular gang if 70% or more of the posts about them tagged the same gang, we were able to reliably measure the gang affiliations of 423 individuals. After limiting the data to gangs with 10 or more affiliates that were verified criminal organizations according to official data, we proceeded with a final dataset of 271 individuals affiliated with 11 gangs (see **Figure 2B**).

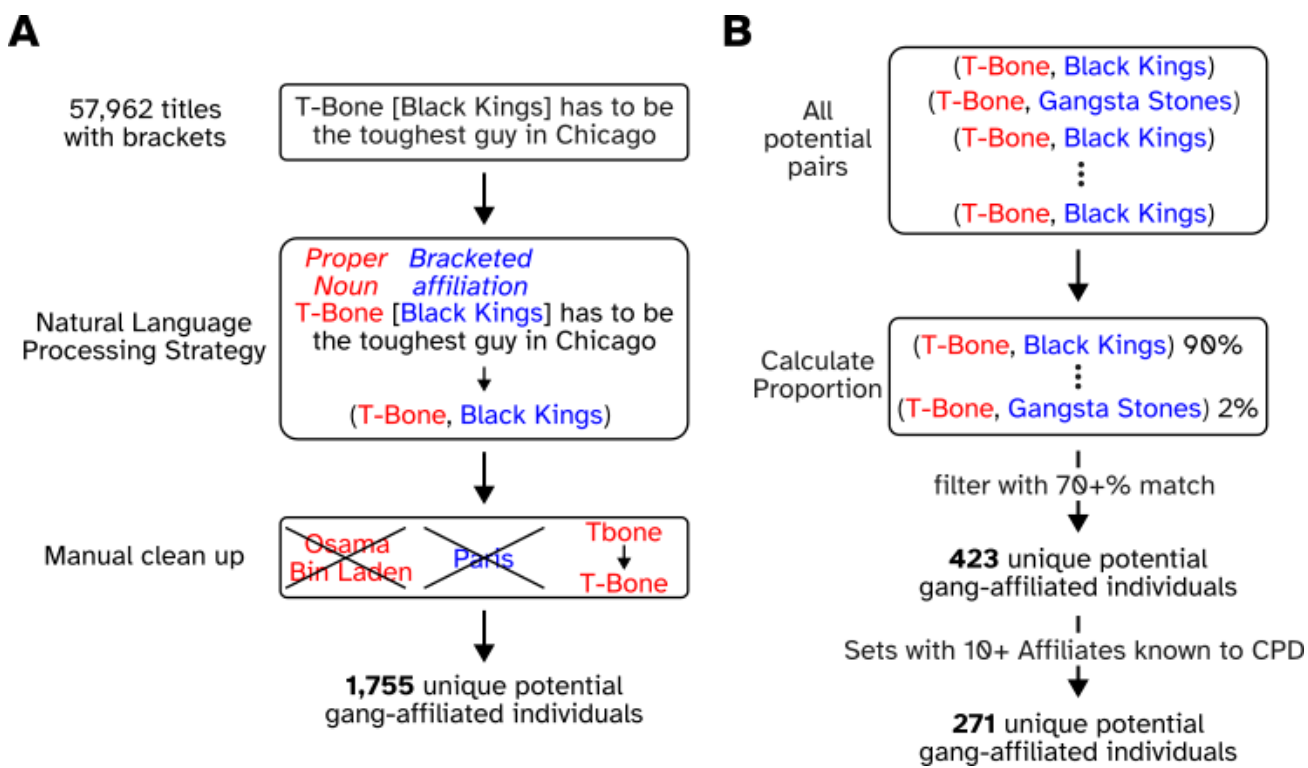

**Figure 2. A.** Methodology for identifying gang-affiliated people discussed by forum users. **B.** Diagram demonstrating processing steps to probabilistically link individuals to the gang they are most likely to affiliate with and generate the final study data.

To build a social network among these individuals, we created edges between individuals who were co-mentioned in at least one post (**Figure 3A**). We were able to identify 1,696 edges from our 271 individuals through our analytic approach. The median individual in our network was connected to 12 other individuals (see **Figure 3C** for distribution of edge counts across individuals). Additionally, we assumed that individuals who are more strongly connected to each other will be co-mentioned in posts together at a higher frequency. As such, edges were weighted based on the number of times that their respective nodes were co-mentioned together. On average, connected individuals were co-mentioned across 5.01 posts**. Figure 3B** presents the weighted network, where nodes are located closer to nodes they are strongly connected to, and colors represent the gang set affiliation associated with each node.

We observed that nodes of the same color tend to be clustered together, broadly suggesting that individuals tended to be more strongly connected to other individuals affiliated with the same gang set.

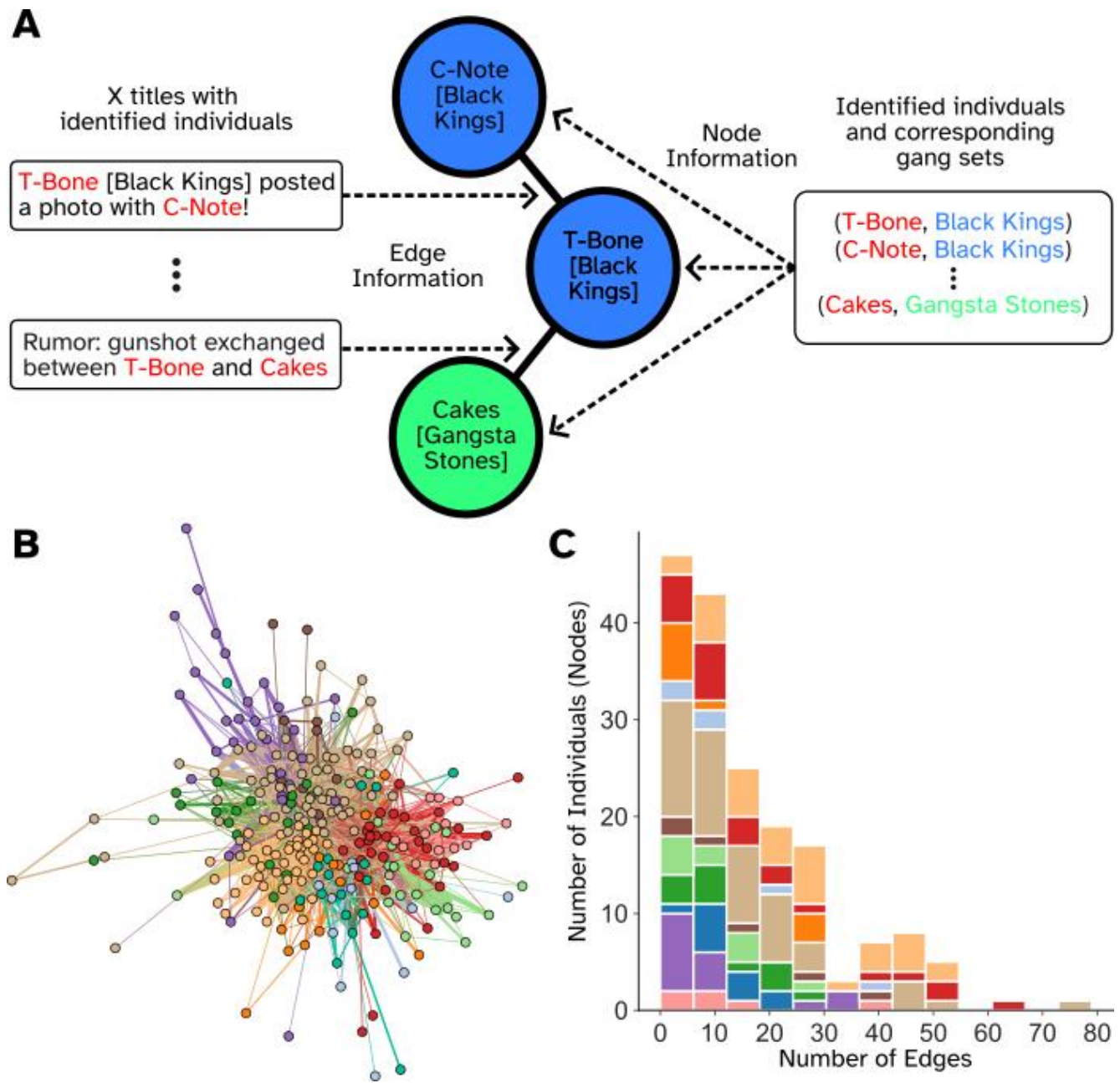

**Figure 3. A.** Methodology for constructing social network from posts containing co-mentions. **B.** Network visualization utilizing the force-directed Yifan Hu algorithm so that nodes tend to appear closer to the nodes they are most frequently co-mentioned with. Each color on the plot represents a unique gang set. **C.** Distribution of edge counts across individuals, with each color representing a unique gang set.

*Social-Media Based Gang Network Reflects the Spatial Proximity of the Gang Sets*

To assess the geographic proximity of the study sets, we draw from a map generated by forum users that reports the locations of 754 Chicago gangs (**Figure 4A**). While the original crowd-sourced map spanned the greater Chicago metropolitan region, the 11 gangs predominantly followed by the forum users were highly concentrated in the Southside of Chicago in an area surrounding Washington Park, with a single set sitting to the south-east of this cluster in the South Shore neighborhood (**Figure 4B**). To validate this crowd-sourced spatial data, we compared the boundaries to arrest locations reported in a CPD dataset that identifies purported gang members while reporting their gang affiliation and location of first arrest, with results suggesting a strong overlap between the two data sources (**Figure 4C**). For 9 out of 11 sets, the crowd-sourced polygon intersected with the police beat with the highest frequency of arrests for that gang, suggesting that the two data sources were highly consistent in spatially locating these gang sets. Assessing the other two sets, one was located just several blocks away from the police beat with the most arrests of affiliated individuals, while the other had no single maximum-arrest police beat, making it impossible to make a one-to-one geographic match across data sources. Overall, these results suggested a high degree of validity across the crowd-sourced polygons.

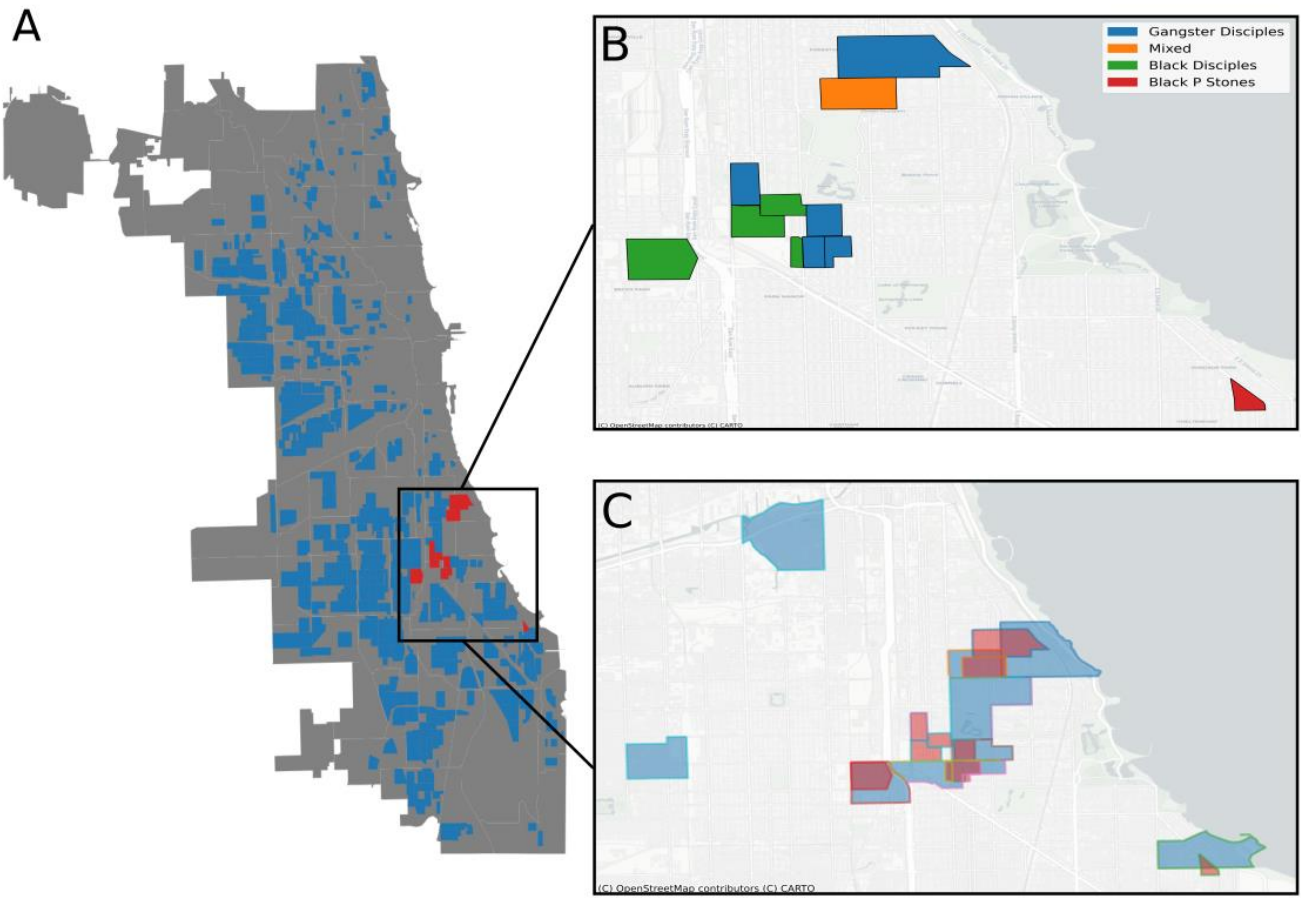

**Figure 4. A.** Map of gang geographies from crowd-sourced map, with analyzed gangs highlighted in red. **B.** Geographies of gangs analyzed in the present study, with colors denoting the gang nation of each set according to CPD data. **C.** Gang set polygons (blue) were laid over the Chicago police beat polygons (red) where affiliates of each gang (outline colors) were most commonly arrested.

Next, we assessed how the crowd-derived social network manifested across gang set turfs. Previously, gang research has highlighted that there tends to be more interaction among gang-affiliated people when their gangs are spatially proximate (Leverso & Hsiao, 2021; Papachristos et al., 2013). Accordingly, the crowd-sourced social network should also show increased connectivity among people from gangs that are spatially closer together. As a first step, we followed the approach of Leverso & Hsiao (2021), who showed that gang-affiliated people were more likely to 'diss' rivals on Facebook when rival gangs were proximate to their home turf. To do so, we created a network capturing the geographic distance in meters between the centroid of each study set. We then created a set-level social network by calculating the average number of times members of a given set were co-mentioned with members of other sets. Consistent with the findings of Leverso and Hsiao (2021), our results indicated that affiliates of different gangs tended to be co-mentioned more frequently when there was smaller distance between sets ($r = -0.22$, $p < .05$).

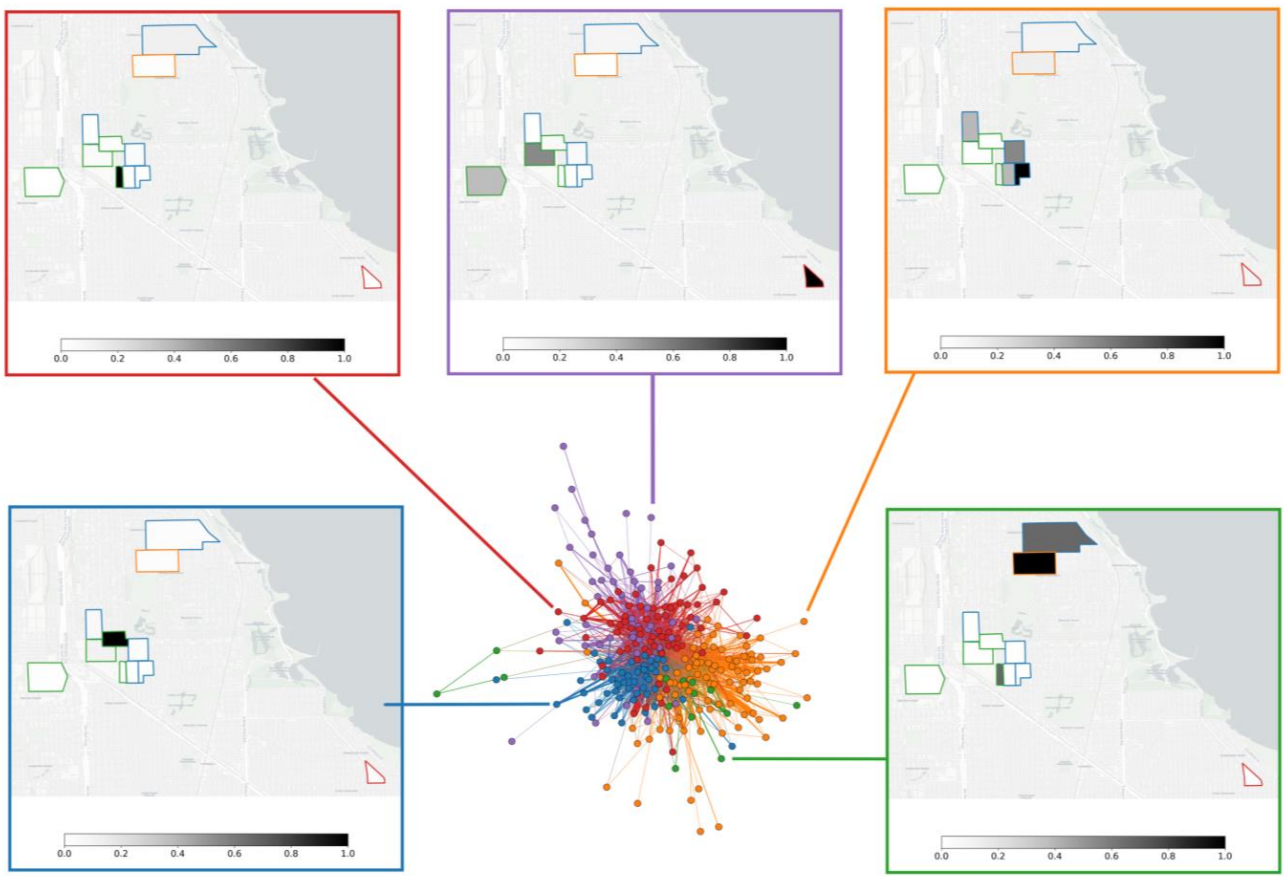

**Figure 5**. Maps of 5 network communities identified within the crowd-sourced gang network. Darker polygons indicate that a given set contains a higher proportion of community members, whereas lighter polygons indicate a lower proportion of community members in the set. Outline color represents nation affiliation.

*Gang-Affiliated People Network with Members of Other Close-By Gangs*
To consider how individual social networks manifested across space and gang sets, we then utilized community detection methods to identify sub-communities within the broader network. Using the Louvain method (Blondel et al., 2008), the nodes and ties in the network of gang-affiliated individuals were analyzed to sort nodes into non-overlapping communities of highly connected individuals. This exercise provided insights about how the social networks of gang-affiliated Southsiders expand beyond fellow members of their own gang, with the results suggesting the crowd-sourced gang network of 11 gang sets was composed of 5 distinct communities. Across these 5 communities, several key patterns emerged (See **Figure 5** for maps of network communities). Visualization of Community 2 (orange) and Community 3 (green) demonstrated spatial clustering of network communities, evidenced by the presence of contiguous polygons with high community representation (darker shades of gray). To confirm this spatial clustering was not a random product of the spatial distribution of nodes in our network, these maps were compared maps of networks randomly simulated from our data (See Supplementary Material). Beyond spatial clustering, the results also suggested that social network communities may form around a shared nation affiliation, as Community 2 was composed of sets that were all affiliated with the same nation. However, the sets associated with Community 3 stem from varying nations, suggesting that the effects of someone's nation affiliation on the composition of their social network may be nuanced and context-dependent.

In contrast to the aforementioned communities, Communities 1 (red) and 4 (blue) were each composed of a single set. Additionally, the geographic patterning diminished for Community 5 (purple), which included two proximate sets west of Washington Park and a third set miles away in the South Shore. This degradation of geographic clustering was not necessarily surprising, as the last group detected by any clustering algorithm will inherently contain outlying nodes that are disconnected from the smaller, more densely connected sub-networks within the broader network. Overall, these results indicated that people in our crowd-sourced network were frequently socially connected to individuals from other gang sets, particularly those affiliated with spatially proximate gangs and nations.

*Individuals' Features in the Social-Media Based Gang Network are associated their Mortality Risk*
Given that previous studies using police-based data have identified that network features correlate with gun victimization, we tested whether features of our crowd-sourced network of gang-affiliated people were predictive of mortality. For these analyses, we collected information from a table maintained by the forum users that recorded deaths from gang violence, allowing us to identify which individuals in our social network were homicide victims. To consider the ways that features of specific gang nations and sets impact mortality risk among their affiliates, we utilized a 3-level hierarchical modeling strategy to nest individuals within their set and nation, allowing us to compare whether features of nations, sets, or individuals were most salient in explaining statistical variation in mortality. To assess the degree of variation in mortality risk explained across this hierarchical structure, we conducted a variance decomposition analysis by evaluating the intercept and variance components of a model with no independent variables included (**Table 1**). With 97.07% of variance in mortality apportioned to the individual-level, these results suggested that mortality risk among gang members was driven by individual characteristics rather than features associated with gang affiliation. Sensitivity analyses demonstrated that accounting for community clusters did not explain additional variation in mortality (See Supplementary Material). While past research in Chicago at the nation-level has demonstrated that gang affiliation puts one at risk for mortality (Papachristos, Braga, et al., 2015), these results indicated that this degree of risk does not seem to change depending on which gang nation or set a person is affiliated with.

|  | Variance Components (Percentage of Variance Across Structure) |
|---|---|
| Individual-level | 3.29 (97.07%) |
| Set-level | 0.04 (1.13%) |
| Nation-level | 0.06 (1.81%) |

Table 1. Variance Decomposition from Empty HLM Model (n=271)

Table 2 reports results from a 3-level regression model that included network features from past research showing network effects of violence, such as an individual's centrality in the network (number of edges) and the percentage of their co-mentions that are with homicide victims. We found that individuals in our network were more likely to be deceased when a higher proportion of their direct neighbors were deceased (O.R. = 2.22, p < .001). Individuals in the study sample who had no direct deceased neighbors had a baseline mortality rate of 21.05%, whereas an individual with 20% of their direct neighbors deceased has an expected mortality rate of 37.18%.

|  | Odds Ratio (Standard Error) |
|---|---|
| Intercept | 0.52 (0.21)** |
| Degree Centrality | 1.90 (0.25)* |
| Percentage of Direct Neighbors Deceased | 2.22 (0.17)*** |
| Edge Weight to Deceased Direct Neighbors | 0.70 (0.25) |
| Percentage of Direct Neighbors Within Gang | 1.29 (0.17) |
|  |  |
| Set-Level Variance Component | 0.00 |
| Nation-Level Variance Component | 0.06 |
| AIC | 330.7 |
| Composite $R^2$ | 0.18 |

Table 2. Logistic Regression Model Assessing Network-Based Correlates of Mortality (n = 271)

Additionally, while past studies have not found a statistically significant association between network centrality and mortality in the police data-based network (see: Papachristos et al., 2012), we found that network degree centrality was strongly associated with mortality in the crowd-sourced network (O.R. = 1.90, p < .05). Given that the typical person in the network has a mortality risk of 35.1% and is connected to 6.54% of the network (17.74 ties), these results suggested that a one-standard-deviation increase in their edges (an additional 18.16 ties) will raise their mortality risk by 90% (66.69%). To verify this result was not the product of a phenomenon where violence victims were systematically discussed more centrally on the forum, ERGM analysis (Jackson, 2011) was used to assess whether victimized individuals had a greater tendency to form network ties, with results suggesting no relationship (See Supplementary Material). Sensitivity analyses also demonstrated that these results were generally robust to a single-level modeling specification as well as varying thresholds for set affiliation linking, and that further variation in mortality was not explained by accounting for individual's centrality within their gang set or nation. (See Supplementary Material).

## DISCUSSION

Results from recent qualitative studies have suggested that there has been a shift in the social dynamics surrounding gangs in Chicago's Southside (Aspholm, 2020; Stuart et al., 2020). Previously, ethnographers have demonstrated that young men from these neighborhoods are increasingly identifying with smaller gangs (Aspholm, 2020), projecting violent gang imagery to online audiences (Stuart, 2020), and largely perpetrating violence that is unrelated to resource competitions between gangs (Aspholm, 2020). However, these changes have been nearly impossible to study through the traditional police-generated datasets that dominate the gang literature. Here, we showed that, by tapping into knowledge bases developed by fans who consume gang-centric social media content, we can generate alternative datasets that augment and improve upon police knowledge. In doing so, we produced the first quantitative evidence that the social networks of gang-affiliated people in Chicago's contemporary Southside are not bound by gang set or nation affiliations and that risk of violent victimization is associated with connections to gang-affiliated people in general, rather than specific gang sets or nations. Consistent with past research showing that violence is contagious within social networks, this study also demonstrates that networks of violent contagion can be captured through NLP-derived, crowd-sourced social media data. We now consider the implications of this work.

*Hyperlocal Gang Affiliations and EBK Dynamics: A Changing Landscape*

By analyzing crowd sourced social media-based data we were able to develop a better quantitative picture of how individuals affiliate with gangs and other gang-affiliated people in the modern-day Chicago Southside. While research on Chicago gangs has generally focused on the traditional nation groups, we observed that forum users and the gang social media content they drew from generally discussed gang affiliations in hyperlocalized set nomenclature. This is consistent with the qualitative results produced by Aspholm (2020), suggesting that Chicago's traditional, large gang nations have splintered into smaller gang set factions with turfs that may span just a few city blocks. In light of these findings, we argue that researchers cannot fully understand gang processes in modern day Chicago without accounting for the set affiliations of involved individuals.

In evaluating how social network connections manifested across these gang sets, we found that individuals in the network were frequently connected to people from other sets. This, again, is consistent with the argument from Aspholm (2020) that gang affiliations are less salient in driving social dynamics among today's young people. While we did find evidence of social communities composed of individuals who all shared a single set affiliation, we also found evidence that there are communities composed of individuals from multiple sets, wherein individuals are more likely to be connected if they are affiliated with sets that are close to one another geographically. Future research should further investigate how gang-affiliated people form social connections with people in their area despite potentially incongruent gang affiliations.

Furthermore, we found that accounting for set and nation affiliations added marginal explanatory power to models predicting mortality. In other words, our results indicated that we do not necessarily need to know an individual's gang affiliations to model their risk of dying from gang violence. This is consistent with Aspholm's (2020) argument that gang violence in the Southside is the product of interpersonal conflicts among "everybody killers" rather than between-group conflicts. Considering this theoretical framework, it is intuitive that features related to an individual's position in the network of gang-affiliated people would be the main drivers of their mortality risk. In this way, Aspholm's EBK argument is congruent with the broader body of research suggesting that gang violence is a contagion that flows through social networks. However, questions still remain about *why* gang-affiliated people perpetrate violence on one another. While Aspholm points to interpersonal conflict among individuals, the perspective of the broader gang literature suggests that violence is the product of feuds between groups. While results from the present study are more suggestive of the former mechanism, future research should work to identify the set affiliations of shooting perpetrators and victims to further investigate whether conflicts between sets play any role in generating gang violence.

*Pushing Gang Research Forward with Crowd-Sourced Social Media Data*

Due to the depth of knowledge about gang-affiliated people that is publicly available through social media, crowd-sourced data can facilitate studies that answer previously inaccessible questions. At the individual scale, crowd-sourced data can be used to understand the people who compose violence networks. While the present study emphasized just one crowd-sourced personal feature, gang set affiliation, there are many other types of information that could be extracted about gang-affiliated individuals based on forum posts. In one promising direction, metadata about how often forum users discuss each individual could be embedded with informative status indicators that are suggestive of victimization risk. Moreover, analysis of the words forum posters use to discuss different gang-affiliated people could provide valuable context to understand the individuals within the network. For example, if a gang-affiliated person publishes a series of videos threatening others and brandishing weapons, forum users may increasingly use words related to violence and guns when they post about that individual.

Additionally, the information embedded in social media posts about gang-affiliated people can also be used to better understand the relationships among individuals within the network. While this study demonstrated that important outcomes can be modeled using a networking strategy that simply connects any two people co-mentioned on the forum, social media data offer cues that could be used to measure these ties with more precision. For example, developing techniques to differentiate social connections between friends and enemies based on social media text content or sentiment would greatly enhance our ability to predict mortality and to understand how conflicts play out within and across gang sets. Beyond basic information about personal relationships, the level of knowledge claimed by forum users is so high that there are countless ways to develop more sophisticated measures of network and node features. For example, posts about brothers and cousins could be analyzed to create familial networks and assess how family structures and neighborhood-based gang structures intertwine to explain which gangs individuals affiliate with. Semantic information posted about individuals over time could be analyzed to see how relationships within the network change over time, which could be useful for answering questions about why people change or leave gangs. While police-based network datasets are generally only able to capture connections among people who are arrested or investigated together, we argue that the social media-based approach presented in this study provides opportunities to research gang networks at a level of specificity and nuance that is impossible through traditional police-generated datasets.

*Conclusion*

While the academic literature on urban street gangs has generally relied on qualitative approaches and quantitative datasets produced by police departments, this study introduced a new approach of using crowd-sourced social media data. Through this approach, we were able to gain new insights into the nature of social relationships among gang-affiliated people in the modern-day Southside of Chicago, with results suggesting that social networks and violence are not as neatly bounded in faction-based dynamics as is often assumed throughout the literature. Beyond gang research, this study also demonstrated that posts by social media users reflect knowledge bases that can be aggregated to create measures bigger than the sums of their parts. While the scientific literature about online spaces has been proliferated with studies measuring features of users (e.g., studies considering how social media usage varies across people from different genders, races, and political parties), results from the present study suggest it can also be fruitful to extract knowledge from users to learn about extrinsic aspects of the social world. Given past research suggesting that it can be scientifically fruitful to extract collective information from swaths of people to develop measures of crowd wisdom (Cooke, 1991; Davis-Stober et al., 2014; Surowiecki, 2005), we propose that computational analysis of online users presents a viable strategy to learn from the crowd. It is well documented that people meet together online for the purpose of discussing shared interests such as musical groups (Bennett, 2012), television shows (Andrejevic, 2008), and even extremely niche subjects such as Hard Rock Cafe pins (Geraghty, 2014) and school shooters (Oksanen et al., 2014). While any given individual is limited in how much accurate knowledge they can have in a given domain, scholars have highlighted that aggregated knowledge among experts can be highly valuable (Fiechter & Kornell, 2021; Goldstein et al., 2014; Mannes et al., 2014). We argue that users of subject-focused online communities such as the one analyzed in this study represent informal expertise that can be leveraged to teach academics about otherwise difficult to access information. Through tapping into the knowledge of online experts, we can easily learn about populations or subjects that have been challenging, if not impossible, to access through traditional quantitative data collection strategies. As such, the usage of crowd-based social data can facilitate new forms of theory testing across social scientific disciplines. The present study exemplified this approach by leveraging posts from gang fans to illuminate the nature and nuances of gangs and cross-gang social networks in Chicago.

## MATERIALS AND METHODS

*Online Data Source*

Specifically, we focused on two Reddit forums started in that tout themselves as places to discuss "gang banging in Chicago" (Following the approach of Stuart and colleagues (2020), to protect the privacy of research subjects we have reworded all quotes so the forum and its users cannot be easily found through Google). The larger forum has over 200,000 subscribed users today, signifying a sizable online community of users posting about gang-affiliated people. In gathering data from these forums, we downloaded 281,871 titles of posts made between January, 2018 and March, 2023. While the earliest forum was launched in 2017, not enough posts were made to analyze posts from that year. These titles were generally about a sentence long and served to contextualize an image, video, or longer text post that other users may respond to. Often, when discussing individuals with perceived gang affiliations, users wrote the alias of a gang-affiliated person and then listed their gang affiliation in brackets or parentheses. **Figure 2** depicts a hypothetical post title where a user tags the individual "T-Bone" as a person affiliated with the Black Kings gang. To isolate posts where gang affiliates were discussed using this syntax, we identified 57,962 post titles that included parentheses or brackets. To extract the aliases of individuals referenced, we used SpaCy's en_core_web_trf model (version 3.5.0) to conduct parts-of-speech tagging and identify proper nouns. After manually removing strings that were highly unlikely to be depicting gang-affiliated people living in Chicago (e.g., locations, famous rappers from other cities, world figures such as Barack Obama and Osama Bin Laden, etc.) and combining strings that depicted the same individual (e.g., "T-Bone", "T Bone", and "Tbone"), we produced a list of 1,755 people discussed on the forum. To capture the gang affiliations assigned in each post, we extracted all strings contained within brackets or parentheses. After removing strings that were unlikely to represent gang labels, this produced a list of 214 possible gangs.

*Identifying Gang Affiliations*

We assumed that individuals were affiliated with the gang that they were most frequently tagged with. Notably, there was high variability in how an individual's affiliation was tagged across posts. If there is ambiguity in the affiliation of a certain individual, forum users may disagree about that individual's affiliation, in which case that person would have several gang tags across posts. Moreover, some gangs may go by multiple names or even change names, creating further instability in how individuals were tagged by the community. As such, we used a probabilistic approach to link gang members to specific gang affiliations. Specifically, we assumed an individual was affiliated with a particular gang if 70% of the posts mentioning that individual tagged the same gang. 70% was chosen as the threshold because it allows expected ambiguity in tags to emerge across users (as opposed to a threshold of 90%+) while screening out individuals whose affiliation has not been established by consensus according to the online community. The methodological strategy allows for ambiguity because users may sometimes inaccurately identify sets or there may not be consensus among all users about the set affiliation of a given individual. Individuals referenced four times or less in total across all posts were removed.

Despite this careful procedure, our NLP approach and the data to which it was applied are inherently unstable. To further ensure our analyses accurately characterized real gang dynamics, we imposed two additional exclusion criteria on our data at the set level. First, we excluded any gangs to whom 10 or fewer affiliates were ascribed membership; although it is possible that authentic and active gangs exist with 10 or fewer members, it is also likely that the gangs we observed with few affiliates are large groups that are simply not reported on by the social media users. Through this approach, we reduced our sample from 59 total gangs with at least one affiliate to 12 gangs with 10 or more tagged affiliates. Second, we compared our list of gangs to those listed in a Chicago Police Department (CPD) dataset of gang-affiliated people (ProPublica, 2018). Despite the already-discussed drawbacks of relying on CPD data, these data allowed us to verify that the groups reported on by the forum are indeed purported criminal organizations. Of 12 gangs with ten or more affiliates, 11 were listed in the CPD records. As such, this study proceeded with a final analytic dataset capturing 11 gangs composed of 271 affiliated individuals.

*Social Network Measurement*

After establishing a dataset of gang-affiliated people, our next step was to generate a network connecting them. In doing so, we set out to mirror the approach from police-based studies, where individuals are linked together when they have been co-mentioned in arrest records or field investigation reports (see: Green et al., 2017; Papachristos et al., 2012; Papachristos & Wildeman, 2014). The present study assigned ties between individuals if they were co-mentioned in at least one social media post. Note that this matching utilized titles of all posts in the message board in the given time period, while the identification of the gang-affiliation of people was done only using titles that contained the bracket structure discussed above.

*Geographic Measurement*

With a list of Chicago gangs and affiliated members in hand, we next set out to identify the geographic boundaries associated with each gang. To do so, we drew from a crowd-sourced map posted on the forum by a user. Using Google Maps tools, the user (or potentially a group of users) made an interactive map where polygons were drawn that indicated the boundaries associated with over 1,000 gang sets in the Chicago metropolitan area. Within each boundary, the forum users identified the name of the gang sets and nations associated with that area. This map was shared as a link through a post on the forum in a locked format, suggesting it had not been freely alterable by uninformed users or "trolls." This source was frequently cited by forum users when engaging in debates related to the geographic location of gangs. Because the map was shared on the forum as a direct link to the Google Maps page, we were able to download the raw spatial data and ingest it into our analyses using the GeoPandas Python library. The crowd-sourced spatial data was manually screened to identify polygon labels that corresponded to the NLP-derived gang labels, with matches being made for all gangs in our sample.

To validate this crowd-sourced spatial data, we compared our boundaries to arrest locations reported in the CPD dataset on gang-affiliated people. For each person CPD identified as being gang affiliated, the dataset reported the person's identified set and the police beat where their first arrest occurred. Using this information, we were able to create maps for each gang that visualize the arrest frequency of affiliated individuals across policing geographies. Assuming that gang-affiliated people were most likely to be arrested closest to the turf associated with their gang, we assessed the validity of our crowd-sourced geographies by evaluating how far each crowd-derived polygon was from the locations where affiliates most frequently get arrested.

*Measuring Mortality*

To identify individuals in our study network who fell victim to gun violence, we leveraged information from a crowd-sourced mortality dataset. Users of the forum of interest maintained a table about gang affiliated people who have died, with the table reporting aliases, government names, and set affiliations for deceased individuals, linking to relevant media reports when possible. This crowd-sourced mortality dataset was downloaded and manually screened to identify referenced aliases that were also present in the study network. In some cases, aliases in the network appeared multiple times in the mortality table. In such cases, we joined records only when the alias and set in our network matched to the mortality network. Cases were dropped where an alias in our network appeared multiple times in the mortality data but there were no matches on set affiliation. Of the 271 analyzable individuals, we identified 95 that died (35.06%). While this mortality rate is high relative to the general population, it is not surprising given past research showing the increased victimization risk associated with participating in gangs (see: Papachristos, Braga, et al., 2015)

*Assessing Correlates of Mortality*

To evaluate if the crowd-sourced data performed comparably to police-derived gang networks, we set out to mirror the approach of past studies that evaluated network effects of mortality (see: Green et al., 2017; Papachristos et al., 2012, 2013). Following these previous approaches, individual-level measures were constructed from our crowd-sourced network. *Degree centrality* was measured as the percentage of nodes with which a given individual shared an edge, a commonly used indicator that captures how central a node is in the network. *Percentage of deceased neighbors* was measured as the percentage of nodes with which a given individual shared an edge who had passed away as of March 2023. *Edge weight to deceased neighbors* was measured as the average number of co-mentions a given individual had among deceased direct network neighbors. Finally, *Percentage of neighbors within gang* is measured as the percentage of an individual's edges to other members of the same gang as a fraction of their total edges.

In response to characteristics of our data, our analytic approach deviates from past studies in several ways. Due to the density of mortality within the analyzed network, the shortest-distance-to-victim measure utilized by past studies, such as Papachristos and colleagues (2012), would equal 1 for the vast majority of subjects and, thus, is not included. Additionally, because all individuals within our network are purportedly in gangs, we replace the percent of network in gang variable used in past work with a measure representing the percentage of each individual's direct network who are affiliated with the same gang.

*Hierarchical Linear Modeling*

Because each individual in the crowd-sourced social network is affiliated with a gang set which is affiliated with a gang nation, individual mortality risk was modeled using a 3-level hierarchical structure. Hierarchical modeling has been frequently used in social research to analyze individuals clustered within contexts, such as in the cases of students in the same classroom or survey respondents living in the same neighborhood (Raudenbush & Bryk, 2002). Using random intercept models, mortality risks are regressed with a structure defined as (Greene 2007):

$N_{ij}$ = individuals within sets
$M_i$ = sets within nations
$L$ = nations
$y_{ijk}$ = mortality rate for individual $k$, set $j$, nation $i$

Using the following equation:

$$logit\left(P(y_{ijk}=1)\right) = \log\left(\frac{P(y_{ijk}=1)}{(y_{ijk}=0)}\right) = X'_{ijk}\beta + u_{ij} + v_i + \varepsilon_{ijk}$$

where:

$X'$ = Independent predictors
$u$ = Set-level error term
$v$ = Nation-level error term
$\varepsilon$ = Individual-level error term

Because this model is estimated as a logistic regression, when conducting variance decomposition, we followed the approach of Goldstein et al. (2002) and Bosker & Snijders (2011) by fixing the intercept to (pi squared over 3).

Crowd-Sourced Gangs 19

# SUPPLEMENTARY MATERIALS

*Randomized Network Clusters do not Geospatially Cluster*

While the contiguous groups of polygons observed in the community cluster maps are suggestive of spatial clustering, it is possible these clusters emerged randomly based on the spatial distribution of nodes in our network. To assess this possibility, we randomized the community cluster label for each node and averaged the cluster representation over 10,000 iterations to visualize the spatial structures of the network communities that would randomly emerge from the crowd-sourced gang network. The randomized maps did not contain contiguous groups of polygons, but instead reflected the sizes of each set, with more populous sets appearing more frequently on average across all clusters. These results suggested the spatial patterns observed in the crowd-sourced gang communities are not a product of random chance.

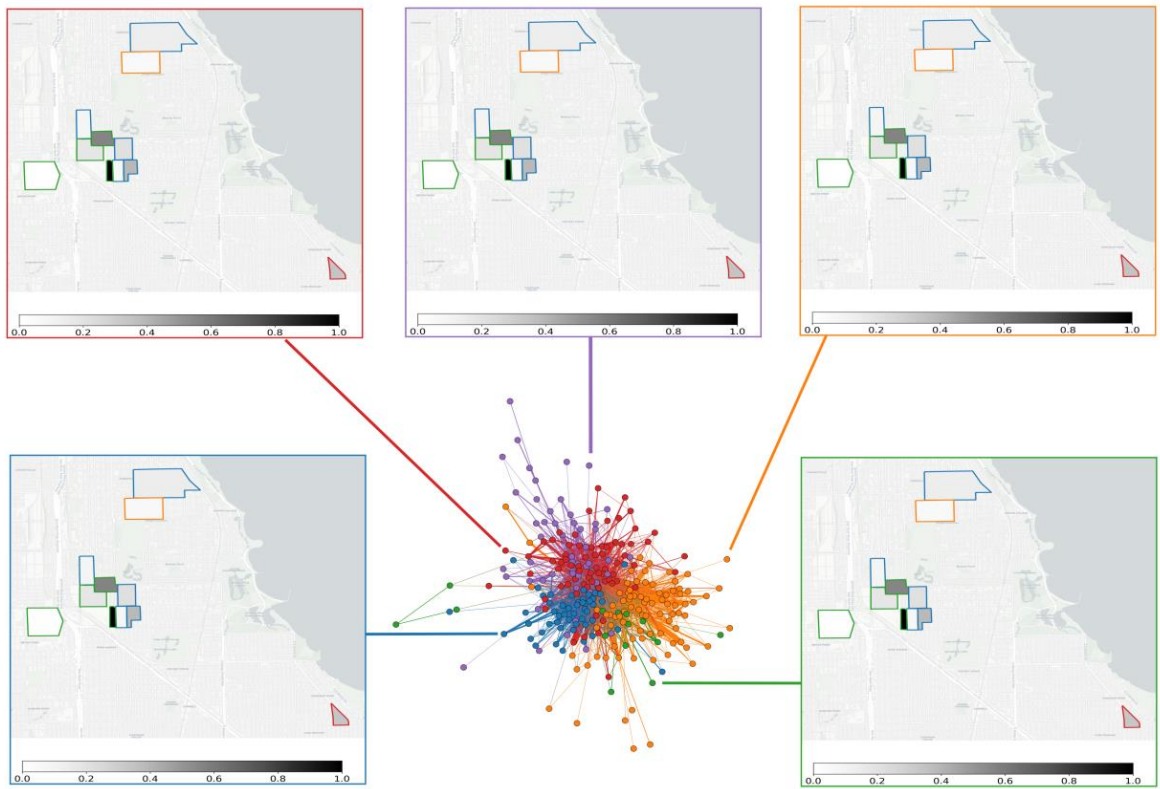

Maps of 5 randomized network communities. Darker polygons indicate that a given set contains a higher proportion of community members, whereas lighter polygons indicate a lower proportion of community members in the set. Outline color represents nation affiliation.

*Mortality Risk Does not Meaningfully Vary Across Network Communities*

We also considered that the social network communities we identified through the Louvain method could represent a third meaningful level of the hierarchical structure. To assess this possibility, we conducted variance decomposition on a 4-level model that nested individuals within sets, nations, and network communities. Results suggested there is no variation in mortality apportioned to network communities.

| | |
|---|---|
| Individual-level | 3.29 (98.55%) |
| Set-level | 0.04 (1.28%) |
| Network Community-level | 0.01 (0.02%) |
| Nation-level | 0.06 (1.75%) |

Variance decomposition from empty HLM model nesting individuals within gang sets, network community, and gang nations (n = 271).

*ERGMs Suggest Mortality Not Driving Centrality*

We also considered the possibility that the association between network centrality and mortality could be attributed to a phenomenon where online posters have a systematic tendency to disproportionally report about deceased individuals. To do so, we utilize exponential random graph models (ERGMs), which describe the probability of a given graph arising as a function of a set of that graph's features. To ask whether the ties formed in our network were meaningfully associated with the prevalence of mortality in it, we used an ERGM to pose our network as a function of individual mortality while holding network-level edge count and degree distribution constant. Results indicated that the likelihood of tie formation in our observed network was not meaningfully influenced by the attribution of mortality to individuals in it.

| | Estimate (std. error) | MCMC |
|---|---|---|
| | 0.01 | 0.02 |
| Null Deviance | 0.00 | |
| Residual Deviance | -0.30 | |
| AIC | 1.70 | |

Exponential random growth model

*Using a Single-Level Bootstrapped Regression Does not Alter Results*

Given the low proportion of variance attributed to the set- and nation-levels of the hierarchical structure, we estimated a single-level regression model predicting mortality using the same parameters. This model that does not account for set or nation affiliation had a comparable AIC (327.29 compared to 328.1) and a substantially lower $R^2$ value (pseudo-$R^2$ of 0.10 compared to composite $R^2$ of 0.18 in the multi-level specification), indicating our model gained explanatory power by accounting for the multi-level nature of gang social structures.

|  | Odds Ratios (Standard Error) |
|---|---|
| Intercept | 0.49 (0.14)*** |
| Degree Centrality | 1.79 (0.23)* |
| Percentage of Direct Neighbors Deceased | 2.27 (0.16)*** |
| Edge Weight to Deceased Direct Neighbors | 0.72 (0.24) |
| Percentage of Direct Neighbors Within Gang | 1.22 (0.15) |
|  |  |
| AIC | 327.9 |
| $R^2$ | 0.10 |

Logistic regression model assessing network-based correlates of mortality using a single-level specification with bootstrapped standard errors (n = 271).

*Altering Affiliation Thresholds does not Substantially Alter Results*

Additionally, we considered that we may have biased our study sample when we removed all individuals that were matched to their gang affiliation in less than 70% of all posts that referenced them. To evaluate how this choice impacted our analytic results, we estimated models on samples generated using 51% and 90% thresholds. While the observed relationship between percentage of network neighbors deceased and mortality was robust across models, the centrality measure was rendered statistically insignificant in the model based on the 90%-threshold sample. The 90%-threshold model had a much smaller sample size (142 individuals) and substantially lower composite $R^2$ value (0.08), suggesting that this version of the model may have lacked the statistical power required to detect a statistically significant relationship for the network centrality parameter. Notably, parameters maintained their directionality across all model specifications.

|  | 51% Threshold (n=365) | 90% Threshold (n=142) |
|---|---|---|
|  | Odds Ratios (Standard Error) | Odds Ratios (Standard Error) |
| Intercept | 0.38 (0.19)*** | 0.48 (0.19)*** |
| Degree Centrality | 1.69 (0.20)** | 1.09 (0.26) |
| Percentage of Direct Neighbors Deceased | 1.80 (0.14)*** | 1.74 (0.20)** |
| Edge Weight to Deceased Direct Neighbors | 0.84 (0.21) | 0.89 (0.26) |
| Percentage of Direct Neighbors Within Gang | 1.25 (0.14) | 1.06 (0.19) |
|  |  |  |
| Set-Level Variance Component | 0.00 | 0.00 |
| Nation-Level Variance Component | 0.06 | 0.00 |
| AIC | 418.5 | 185.8 |
| Composite $R^2$ | 0.13 | 0.08 |

Logistic regression model assessing network-based correlates of mortality at different affiliation thresholds

*Within-Gang Centrality Measures do not Explain Mortality Risk*

We also evaluated whether individuals face additional mortality risk if they have a high degree of social centrality within their set or nation. By creating discrete networks for each set and nation and calculating centrality scores for all nodes, we estimated models including two new parameters: within-set centrality and within-nation centrality. Results suggested these new centrality measures did not help to explain variation in mortality and their inclusion led to a moderately higher degree of over-fitting.

|  | Odds Ratios (Standard Error) |
|---|---|
| Intercept | 0.54 (0.26)* |
| Degree Centrality (Global) | 2.19 (0.44) |
| Percentage of Direct Neighbors Deceased | 2.21 (0.17)*** |
| Edge Weight to Deceased Direct Neighbors | 0.70 (0.25) |
| Percentage of Direct Neighbors Within Gang | 1.33 (0.18) |
| Degree Centrality (Set) | 0.87 (0.26) |
| Degree Centrality (Nation) | 0.98 (0.32) |
|  |  |
| Set-Level Variance Component | 0.00 |
| Nation-Level Variance Component | 0.02 |
| AIC | 334.3 |
| Composite $R^2$ | 0.19 |
| Logistic regression model assessing network-based correlates of mortality (n = 271). | |